\documentclass{emulateapj}

\usepackage{color}
\usepackage{hyperref}

\newcommand{\name}{SDSS0533}
\newcommand{\othername}{W1906}

\begin{document}
\title{ASASSN-16ae: A Powerful White-Light Flare on an Early-L Dwarf}
\shorttitle{A Powerful Flare on an L Dwarf}

\author{{Sarah J. Schmidt}\altaffilmark{*,1}, {Benjamin J. Shappee}\altaffilmark{2,3}, {Jonathan Gagn\'e}\altaffilmark{4,5}, {K. Z. Stanek}\altaffilmark{6,7},  {Jos\'e L. Prieto}\altaffilmark{8,9}, 
{Thomas~W.-S.~Holoien}\altaffilmark{6,7}, {C. S. Kochanek}\altaffilmark{6,7}, {Laura Chomiuk}\altaffilmark{10}, {Subo Dong}\altaffilmark{11}, {Mark Seibert}\altaffilmark{2}, {Jay Strader}\altaffilmark{10}}

\altaffiltext{*}{sjschmidt@aip.de}
\altaffiltext{1}{Leibniz-Institute for Astrophysics Potsdam (AIP), An der Sternwarte 16, D-14482, Potsdam, Germany}
\altaffiltext{2}{Carnegie Observatories, 813 Santa Barbara Street, Pasadena, CA 91101, USA}
\altaffiltext{3}{Hubble Fellow}
\altaffiltext{4}{Department of Terrestrial Magnetism, Carnegie Institution of Washington, Washington, DC~20015, USA}
\altaffiltext{5}{Sagan Fellow}
\altaffiltext{6}{Department of Astronomy, Ohio State University, 140 West 18th Avenue, Columbus, OH 43210, USA}
\altaffiltext{7}{Center for Cosmology and AstroParticle Physics, The Ohio State University, 191 W.\ Woodruff Avenue, Columbus, OH 43210, USA}
\altaffiltext{8}{N\'ucleo de Astronom\'ia de la Facultad de Ingenier\'a, Universidad Diego Portales, Av. Ej\'ercito 441, Santiago, Chile}
\altaffiltext{9}{Millennium Institute of Astrophysics, Santiago, Chile}
\altaffiltext{10}{Department of Physics and Astronomy, Michigan State University, East Lansing, MI 48824, USA}
\altaffiltext{11}{Kavli Institute for Astronomy and Astrophysics, Peking University, Yi He Yuan Road 5, Hai Dian District, Beijing 100871, China}

\begin{abstract} 
We report the discovery and classification of SDSS~J053341.43+001434.1 (\name), an early-L dwarf first discovered during a powerful $\Delta V < -11$ magnitude flare observed as part of the ASAS-SN survey. Optical and infrared spectroscopy indicate a spectral type of L0 with strong H$\alpha$ emission and a blue NIR spectral slope. Combining the photometric distance, proper motion, and radial velocity of \name\ yields three-dimensional velocities of $(U,V,W)=(14\pm13,-35\pm14,-94\pm22)$~km~s$^{-1}$, indicating that it is most likely part of the thick disk population and probably old. The three detections of \name\ obtained during the flare are consistent with a total $V$-band flare energy of at least $4.9\times10^{33}$~ergs (corresponding to a total thermal energy of at least $E_{\rm tot}>3.7\times10^{34}$~erg), placing it among the strongest detected M dwarf flares. The presence of this powerful flare on an old L0 dwarf may indicate that stellar-type magnetic activity persists down to the end of the main sequence and on older ML transition  dwarfs.
\end{abstract}

\keywords{brown dwarfs --- stars: chromospheres --- stars: flare --- stars: individual(SDSS J053341.43+001413.1) --- stars: low-mass}

\section{Introduction}
\label{sec:intro}
Magnetic activity, as traced by chromospheric H$\alpha$ emission, is ubiquitous at the transition between M and L spectral types (M7--L3; ML dwarfs). In more massive stars, this quiescent activity is often accompanied by flares, but observations of flares on ML dwarfs are relatively sparse. The flares that have been observed on late-M dwarfs are frequently dramatic, including those found in serendipitous spectroscopy \citep[e.g.,][]{Liebert1999,Fuhrmeister2004} and photometry \citep[e.g.,][]{Rockenfeller2006,Schmidt2014} in addition to some dedicated monitoring \citep[e.g.,][]{Stelzer2006,Hilton2011phd}. While late-M dwarfs are significantly cooler than the earlier-M dwarfs and FGK stars where other flares were observed, observations have been consistent with a generally similar energy budget, with optical emission including both a thermal continuum and atomic (both line and continuum) emission \citep{Schmidt2007}.

Initial observations of L dwarf flares did not include the thermal white-light emission that is the hallmark of stellar flares, but were instead limited to sudden elevations in the H$\alpha$ emission without a thermal continuum \citep[e.g.,][]{Hall2002,Liebert2003}. White-light flares have previously been detected on only one early-L dwarf: \citet{Gizis2013} found 21 flares in one quarter of Kepler short-cadence data from the L1 dwarf WISEP~J190648.47+401106.8 (hereafter \othername). Additional detections of white-light flares are necessary to understand the changes in both the underlying magnetic dynamo and the interaction between the magnetic fields and the surface as we examine lower-mass, cooler objects. 

On UT 2016 January 10, the ASAS-SN survey \citep{Shappee2014} detected ASASSN-16ae \citep{Shappee2016}, a $\Delta V\sim-11$ magnitude event on SDSS~J053341.43+001434.1 (hereafter \name), a source with ML dwarf colors. The detection of a flare this strong on a very low mass star is not only further evidence that strong magnetic activity extends to ultracool dwarfs \citep[see, e.g.,][]{Rodriguez-Barrera2015,Pineda2016}, but could have strong implications for the habitability of exoplanets found around ML dwarf hosts \citep{Segura2010}. In this Letter, we characterize the L dwarf source (Section~\ref{sec:data}) and the flare (Section~\ref{sec:flare}) to place \name\ in the context of magnetic activity on ML dwarfs.

\section{Characterizing \name}
\label{sec:data}
We combine the ASAS-SN detections with survey data and additional observations to characterize the flare event and the quiescent properties of \name. 

\subsection{$BVI$ photometry}
The ASAS-SN telescopes observe the entire sky roughly every other day. However, approximately half the sky (including \name) is observed at a slightly higher cadence because it is in the overlap region between multiple fields. As a result, there are $V$-band images taken $\sim2.9$ hours before the flare, during the flare, and $\sim22$ hours after the flare. The ASAS-SN fields were reduced using the standard ASAS-SN pipeline \citep[][in prep.]{Shappee2016b}. No source was detected in the two individual pre-flare images\footnote{A third dithered image was taken, but \name\ was $\sim 25 \arcsec$ off the detector.} or the three individual post-flare images, so we combined each set of images.  

\begin{figure}
\includegraphics[width=\linewidth]{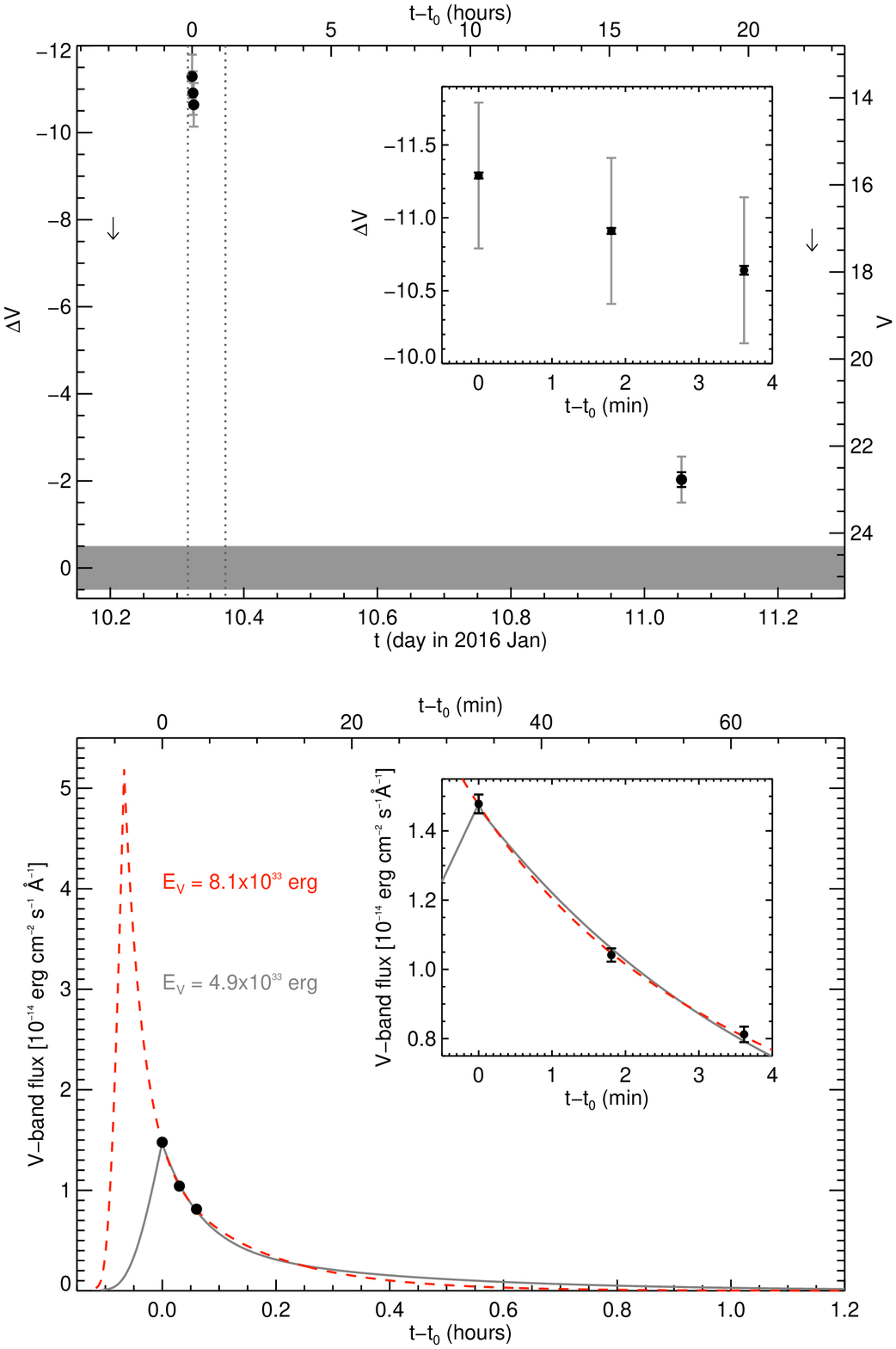} 
\caption{$V$-band detections of \name. Top: $\Delta V$ (left axis) and $V$ (right axis) as a function of time. The quiescent magnitude is shown with its uncertainty (gray shading) and individual detections are shown with photometric uncertainties (black error bars) and uncertainties in $\Delta V$ (gray error bars). The gray dotted line indicates the time included in the bottom panel. Bottom: $V$-band flux as a function of time with the minimum (gray solid) and best-fit (red dashed) empirical flare models (see Section~\ref{sec:flshape}). The inset panels show $\Delta V$ (top) and $V$-band flux (bottom) of the three flare detections.} \label{fig:lc}
\end{figure}

We performed aperture photometry at the location of ASASSN-16ae using the IRAF {\tt apphot} package and calibrated the results using the AAVSO Photometric All-Sky Survey \citep{Henden2014}. We place $3\sigma$ upper limits on the pre- and post-flare epochs of $V>16.7$ and $V>17.0$ respectively, and find a sharp decline from $V=13.5$ to $V=14.16$ during the three individual flare images (see Figure~\ref{fig:lc} and Table~\ref{tab:phot}). ASAS-SN detected no other flares from \name\ in $\sim1100$ images acquired between 2012 January and 2016 March with $V$-band limiting magnitudes ranging from $\sim15$ to $18$ depending on lunation and conditions. 

\begin{deluxetable*}{llllcccc} \tablewidth{0pt} \tabletypesize{\scriptsize}
\tablecaption{Photometric Observations \label{tab:phot}}
\tablehead{ \colhead{Date (UT)}  & \colhead{Source}  & \colhead{$N_{\rm exp}$} & \colhead{ET (s)}  & \colhead{$B$}  & \colhead{$V$}  & \colhead{$I$} & \colhead{$F_V$} } 
\startdata
2016 Jan 10 04:54:08 & ASAS-SN/Brutus & 2 & 90 & \nodata & $>16.74$ & \nodata & $<$7.55$\times10^{-16}$ \\
2016 Jan 10 07:44:22 & ASAS-SN/Brutus & 1 & 90 & \nodata & $13.51\pm0.02$ & \nodata & ($1.48\pm0.03$)$\times10^{-14}$ \\
2016 Jan 10 07:46:10 & ASAS-SN/Brutus & 1 & 90 & \nodata & $13.89\pm0.02$ & \nodata & ($1.04\pm0.02$)$\times10^{-14}$ \\
2016 Jan 10 07:47:58 & ASAS-SN/Brutus & 1 & 90 & \nodata & $14.16\pm0.03$ & \nodata & ($8.12\pm0.22$)$\times10^{-15}$ \\
2016 Jan 11 01:16:04 & duPont & 1 & 60 & \nodata  & \nodata & $19.15\pm0.03$ & \nodata \\
2016 Jan 11 01:19:55 & duPont & 1 & 180 & \nodata & $22.77\pm0.17$ & \nodata & ($2.92\pm0.46$)$\times10^{-18}$ \\
2016 Jan 11 01:25:04 & duPont & 1 & 180 & $23.31\pm0.20$ & \nodata & \nodata & \nodata \\
2016 Jan 11 06:01:31 & ASAS-SN/Cassius & 3 & 90 & \nodata & $>17.01$ & \nodata & $<$5.88$\times10^{-16}$ \\
2016 Mar 29 00:34:32 & SOAR & 1 & 120 & \nodata  & \nodata & $19.43\pm0.03$ & \nodata \\
2016 Mar 29 00:37:20 & SOAR & 2 & 180 & $>24.54$ & \nodata & \nodata & \nodata \\
2016 Mar 29 00:47:20 & SOAR & 1 & 360 & \nodata & $>24.19$ & \nodata & $<$7.90$\times10^{-19}$ \\
\hline
\multicolumn{4}{c}{adopted quiescent} & \nodata & $24.80\pm0.50$ & $19.43\pm0.03$ & ($4.51\pm2.07$)$\times10^{-19}$
\enddata
\end{deluxetable*}

We obtained two additional epochs of $BVI$ photometry from the Wide Field Reimaging CCD Camera (WFCCD) on the duPont 100-inch telescope and the Goodman spectrograph \citep{Clemens2004} on the Southern Astrophysical Research (SOAR) telescope. We performed aperture photometry on these images using the IRAF {\tt apphot} package and calibrated the magnitudes using SDSS Data Release 10 \citep[DR10;][]{Ahn2014} photometry of nearby fields stars transformed into Bessel filters.\footnote{Transformations from \url{http://www.sdss.org/dr12/algorithms/sdssUBVRITransform/\#Lupton2005}} The source was detected on UT 2016 January 11 ($V=22.77$), but the upper limit ($V<24.19$) on UT 2016 March 29 indicates that the January~11 data were taken during a flare. It is unlikely that this detection was part of the ASASSN-16ae event because the largest stellar flares typically last $\lesssim$12~hours \citep{Kowalski2010}. It is possible they are related, however, because a very large flares often  trigger sympathetic flare events \citep{Davenport2014a}. The magnitudes and $3\sigma$ upper limits are given in Table~\ref{tab:phot}.

\subsection{Survey Photometry}
\label{sec:phot}
We obtained PSF photometry for \name\ from the SDSS DR10 \citep{Ahn2014} database as well as from the Two Micron All-Sky Survey \citep[2MASS;][]{Skrutskie2006} and the \textit{Wide-field Infrared Survey Explorer} \citep[\textit{WISE};][]{Wright2010}. The SDSS $iz$, 2MASS $JHK_S$ and \textit{WISE} $W1W2$ are good quality, but the SDSS $ug$ and \textit{WISE} $W3W4$ magnitudes are too faint to be reliable and are not included. The SDSS $r$ magnitude is flagged to indicate the SDSS position is not accurate. We did not correct the survey photometry (given in Table~\ref{tab:prop}) for extinction because \name\ is in the foreground of dust clouds (see Section~\ref{sec:stage}) and not likely to be heavily extincted.

\begin{deluxetable}{lc} \tablewidth{0pt} \tabletypesize{\scriptsize}
\tablecaption{Properties of \name\ \label{tab:prop} }
\tablehead{ \colhead{Parameter}  & \colhead{Value}  } 
\startdata
\multicolumn{2}{c}{Photometric} \\
\hline
$V$\tablenotemark{a} & $24.80 \pm 0.50$\\
$I$ & $19.43 \pm 0.03$ \\
$r$\tablenotemark{b}& $22.44\pm0.15$\\
$i$ & $20.75\pm0.05$ \\
$z$ & $18.87\pm0.04$ \\
$J$ &  $16.38\pm0.13$  \\ 
$H$ & $16.15\pm0.16$\\ 
$K_S$ & $15.47\pm0.21$\\ 
$W1$ & $15.16\pm0.04$\\
$W2$ & $15.07\pm0.08$ \\
\hline
\multicolumn{2}{c}{Spectroscopic} \\
\hline
ST & L0 \\
H$\alpha$ EW (\AA) & $29.5\pm0.5$ \\
$\log(L_{H\alpha}/L_{bol})$\tablenotemark{c} & $-4.1$\\
 $V_{\rm rad}$  [km s$^{-1}$] & $67.3\pm7.0$  \\
\hline
\multicolumn{2}{c}{Derived} \\ 
\hline
d ($M_i$/$i-z$) [pc] & 80.3$\pm$20.7 \\
d ($M_i$/$i-J$) [pc] & 68.9$\pm$22.1 \\
d ($M_i$/$i-K_S$) [pc] & 100.1$\pm$35.2 \\
d ($M_J$/ST) [pc] & 89.4$\pm$17.3 \\
d ($M_H$/ST) [pc] & 115.2$\pm$22.8 \\
d ($M_{K_S}$/ST) [pc] & 103.2$\pm$22.7 \\
d ($M_{W1}$/ST) [pc] & 103.5$\pm$18.7 \\
d ($M_{W2}$/ST) [pc] & 111.6$\pm$20.4 \\
d (mean) [pc] & 96.5$\pm$23.1 \\
$Z$ [pc] & $-$13.3$\pm$6.8 \\
$\mu_{\alpha}$ [mas yr$^{-1}$] & $-124.8\pm$13.6 \\
$\mu_{\delta}$ [mas yr$^{-1}$] & $-156.6\pm$35.6 \\
$V_{tan}$ [km s$^{-1}$] & $-114.6\pm$31.5 \\
$U$ [km s$^{-1}$] & 13.5$\pm$13.4 \\
$V$ [km s$^{-1}$] & $-$35.0$\pm$14.3 \\
$W$ [km s$^{-1}$] & $-$94.0$\pm$21.5 
\enddata
\tablenotetext{a}{Adopted; see Section~\ref{sec:flare}}
\tablenotetext{b}{Centroid flag set}
\tablenotetext{c}{Using $\chi=1.9\times10^{-6}$~\AA$^{-1}$ from \citet{Schmidt2014} }\\
\end{deluxetable}

\subsection{Spectroscopic Observations}
\begin{figure}
\includegraphics[width=\linewidth]{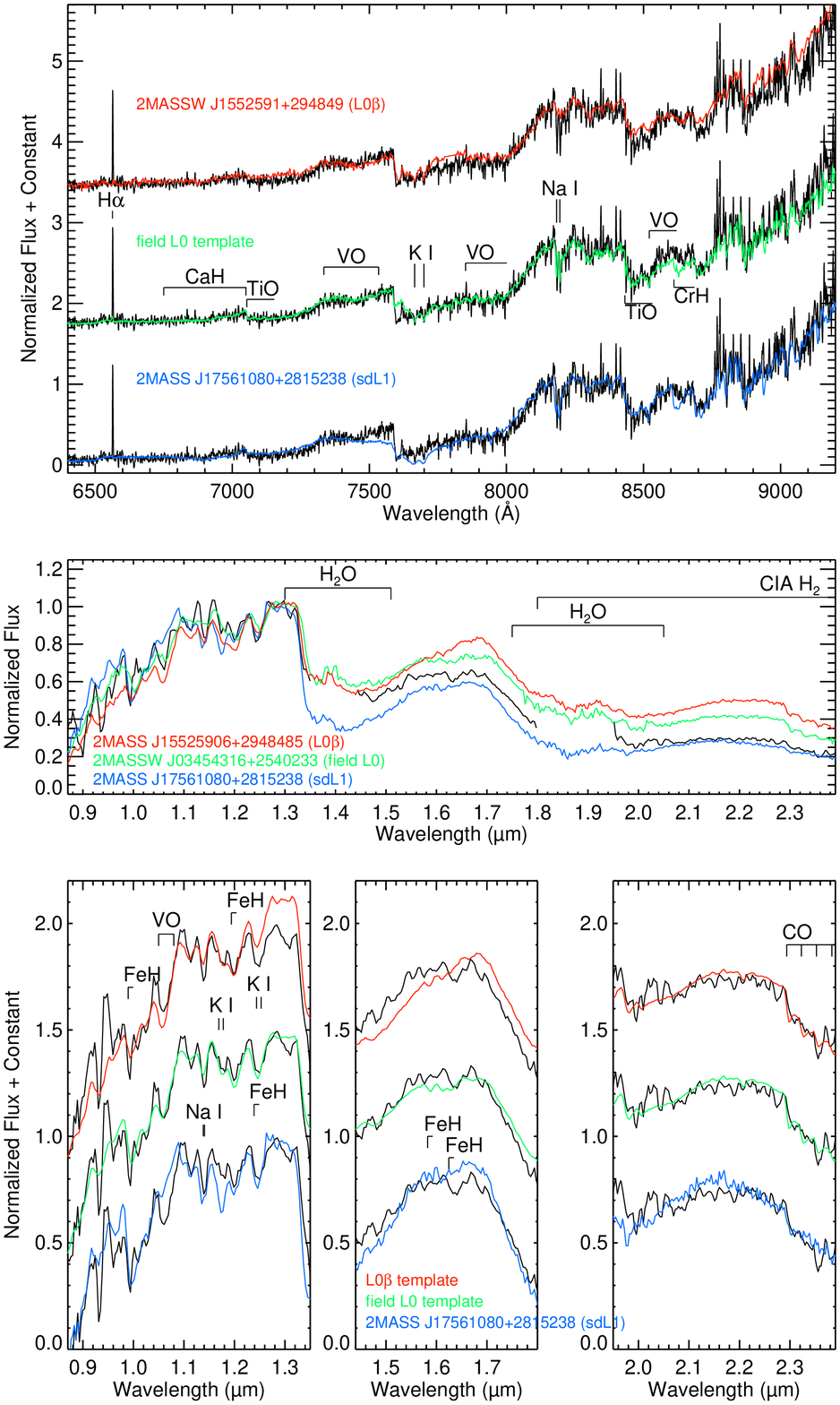} 
\caption{Spectra of \name\ compared to templates, standards, and well-characterized L0/L1 dwarfs. The labeled spectral features were identified by \citet{Kirkpatrick1999} and \citet{Cushing2005}, and the comparison spectra are discussed by \citet{Kirkpatrick2010}, \citet{Schmidt2014a}, and \citet{Cruz2016}. Top: the MagE optical spectrum (normalized at 8150\AA; black) compared to a low-gravity L0$\beta$ standard \citep[red;][]{Reid2008}, the field L0 template \citep[green;][]{Schmidt2014a}, and an sdL1 spectrum \citep[blue;][]{Kirkpatrick2010}. Middle: the FIRE NIR spectrum (black) compared to a low-gravity L0$\beta$ standard \citep[red;][]{Allers2013}, a field L1 standard \citep[green;][]{Burgasser2006a}, and an sdL1 spectrum (blue). The spectra were normalized at 1.28$\mu$m to facilitate comparison of the continuum slope. Bottom row: The FIRE NIR spectrum of \name\ (black) compared to the \citet{Cruz2016} low-gravity (L1$\beta$; red) and field (L1; green) templates and an sdL1 spectrum. The $zJ$, $H$, and $K$ bands are individually normalized to the full range displayed in each panel to facilitate comparison of individual features.} \label{fig:spec}
\end{figure}

We obtained a medium-resolution ($R\sim4800$) optical spectrum of \name\ using the Magellan Echellette \citep[MagE;][]{Marshall2008} spectrograph on the Baade-Magellan (6.5-m) telescope on 2016 January 16. The target was observed for three 1800~s exposures at an airmass of $\sim$1.16 under clear conditions with the 0\farcs85 slit and 0\farcs6--0\farcs7 seeing. The wavelength coverage of MagE is 3100--10000\AA, but the continuum was not detected at wavelengths bluer than $\sim$6200\AA. We extracted all echellette orders of both \name\ and the flux standard LTT~3168 using the MagE pipeline\footnote{\url{http://code.obs.carnegiescience.edu/mage-pipeline}} and flux calibrated the spectra with standard routines from the IRAF {\tt Echelle} package. The wavelength solution was shifted by 0.23\AA\ based on the position of the [O I] 5577 telluric emission line, and a heliocentric velocity correction was applied. The MagE spectrum (shown in Figure~\ref{fig:spec}) is consistent with an L0 spectral type. 

The H$\alpha$ emission line is present in two spectral orders of MagE data. We measure the equivalent width (EW) of H$\alpha$ from both orders, obtaining ${\rm EW}=-29.3$~\AA\  and ${\rm EW}=-29.8$~\AA. We adopt the mean value and assign an uncertainty of 0.5~\AA\  based on the range of the two observations. The EW corresponds to an activity strength of $\log(L_{H\alpha}/L{\rm bol})=-4.1$, one order of magnitude stronger than the average value for L0 dwarfs \citep{Schmidt2015}. 

We obtained near-infrared (NIR) spectroscopy for \name\ with the Folded-port InfraRed Echellette \citep[FIRE;][]{Simcoe2013} at the Magellan Telescopes on UT 2016 January 22 (PI: Gagn\'e). We used the high-throughput long-slit mode to obtain $R\sim450$ across the 0.8--2.45\,$\mu$m range. \name\ was observed in eight 60\,s exposures in an ABBA pattern at an airmass of 1.18. A0 HD~290958 was observed for telluric correction immediately after at a similar airmass (1.17). Standard calibrations were taken as described by \citet{Gagne2015}. The data were reduced using an updated version of the IDL FIREHOSE pipeline, which was based on the MASE \citep{Bochanski2009} and SpeXTool \citep{Vacca2003,Cushing2004} packages \citep[FireHose~2.0\footnote{Available at \url{https://github.com/jgagneastro/FireHose\_v2/tree/v2.0}}; see][]{Gagne2015x}. The spectral slopes of FIRE data are often unreliable due to de-centering of bright A0 standards to avoid saturation, so the slope of \name\ was corrected to match its 2MASS photometry \citep[see][submitted to ApJS]{Gagne2016}. The final NIR spectrum is shown in Figure~\ref{fig:spec}. Spectral indices \citep{Allers2013} and comparison to spectroscopic templates \citep[][submitted]{Cruz2016} indicate an L0 spectral type. 

\subsection{Kinematics}
We estimated the distance to \name\ using the photometric relations of \citet[][in prep.]{Schmidt2016b} and the spectroscopic relations of \citet{Dupuy2012} assuming a spectral type of L0. The individual distance estimates vary from 68.9~pc to 111.6~pc (see Table~\ref{tab:prop}), reflecting the intrinsic scatter of the relations as well as the slightly peculiar colors of \name. We adopt the mean distance of $d=96.5\pm23.1$~pc. We calculated a proper motion of $\mu_{\alpha}=-124.8\pm13.6$ and $\mu_{\delta}=-156.6\pm35.6$~mas~yr$^{-1}$ for \name\ based on the 2MASS and AllWISE coordinates (the SDSS position was excluded due to poor centroiding). A search of the combined SDSS photometric database and USNO-B proper motions \citep{Munn2004} reveals no nearby candidate companions. Combining the proper motion and the distance, we calculate a tangential velocity of $V_{\rm tan}=-114.6\pm31.5$~km~s$^{-1}$, well above the mean for early-L dwarfs \citep[$V_{\rm tan}=28$~km~s$^{-1}$;][]{Schmidt2010}. 

We measured the radial velocity of \name\ from the MagE optical spectrum via cross-correlation with the SDSS L0 template \citep{Schmidt2014a} using five different $\sim500$ pixel regions in the sixth and seventh orders encompassing the region from 7150--8450\AA\ (avoiding telluric absorption at 7600\AA). The derived radial velocity, based on the mean and standard deviation of those values, is $V_{\rm rad} = 67.3\pm7.0$~km~s$^{-1}$. We calculated three dimensional velocities of $(U,V,W)=(14\pm13,-35\pm14,-94\pm22)$~km~s$^{-1}$, with   uncertainties estimated using a Monte Carlo approach, assuming normal distributions for the errors in distance, $\mu$, and $v_{\rm rad}$. The resulting kinematics (given in Table~\ref{tab:prop}) include an unusually high $W$ velocity. Compared to the $W$ distributions of L dwarfs \citep{Schmidt2010}, \name\ is $\sim$5 standard deviations from the mean value for the thin disk, implying thick disk membership.
 

\subsection{The Age of \name}
\label{sec:stage}
\name\ is located in the Orion constellation, within a few degrees of its vast star formation region \citep{Genzel1989}. An association with Orion would imply a very young ($\lesssim$20~Myr) age, but at a distance of $d=97$~pc (compared to $d\sim400$~pc), \name\ is more likely to be a foreground object and so neither within the dust cloud nor part of a young association. L dwarfs with ages $\lesssim$200~Myr can be also be identified through spectral signatures of low gravity from their still contracting atmospheres; the optical and NIR spectra of \name\ are not consistent with low-gravity (L0$\beta$) spectra, instead showing alkali lines, TiO bands, and a spectral slope similar to field-age L dwarf standards \citep[][submitted]{Cruz2007,Cruz2016}. Spectral indices designed to identify low-gravity objects also indicate \name\ is not young \citep[gravity score of 0?00;][]{Allers2013}. 

While young L dwarfs often have red $J-K_S$ colors, \name\ is blue \citep[$J-K_S=0.91$, compared to the L0 median of $J-K_S=1.20$;][]{Schmidt2015}. Bluer $J-K_S$ colors are often associated with older ages, either due to a clearing of dust clouds, lower metallicity, or a combination of those two effects \citep{Burgasser2008a,Kirkpatrick2014}. \name\ has a similar NIR spectral slope to the low-metallicity sdL1 dwarf 2MASS~J17561080+2815238 \citep{Kirkpatrick2010}, but lacks the unusually strong alkali lines and VO bands that are used to identify subdwarfs. Given its large $W$ velocity, \name\ is most likely an older L dwarf (and consequently a star rather than a brown dwarf) with a near-solar metallicity and thin or patchy dust clouds.

\section{Properties of the Flare}
\label{sec:flare}
The properties of the flare are sensitive to both the distance to \name\ and its quiescent $V$-band magnitude. Our $V$-band photometry included only additional flares and upper limits, so we estimated a quiescent $V$-band magnitude based on photometry from \citet{Dieterich2014}. M9.5--L0.5 dwarfs have a mean $M_V=19.9\pm0.4$, $V-J=8.2\pm0.2$, and $V-K_S=9.5\pm0.2$. Based on the distance and $JK_S$ magnitudes of \name, we calculate $V=24.8\pm0.5$, indicating an observed flare amplitude of $\Delta V=11.3\pm0.5$ mag. 

\subsection{Fitting a Flare Shape}
\label{sec:flshape}
To estimate the shape of the flare, we employ the \citet{Davenport2014a} empirical flare model based on Kepler short-cadence observations of the M4 dwarf GJ~1243. This empirical model was built using data from flares that occurred on a much warmer object ($\sim$900~K warmer) and so it is possible that ASASSN-16ae has fundamentally different energetics than the GJ~1243 flares. However, the only spectra of ML dwarf white-light flares \citep[e.g.,][]{Schmidt2007,Gizis2013} are broadly consistent with those from M dwarfs \citep{Kowalski2013}, indicating a flare model from an M4 dwarf is suitable for estimating a flare energy. 

The \citet{Davenport2014a} model uses a double exponential to capture both the initial impulsive and the long gradual decay phases of a  average flare that incorporates 885 classical (single-peaked) flares on GJ~1243. While the Kepler band samples a wider portion of the flare energy distribution than the $V$-band, both bands respond primarily to thermal flare emission \citep[rather than atomic line or Balmer continuum emission; e.g.][]{Kowalski2013}, so their temporal evolution should be similar. The \citet{Davenport2014a} model parameterizes flares based on the amplitude of the peak (here, peak $F_V$) and the time to half-light decay ($t_{1/2}$). Because our data only captures the decay phase, we also fit the time between the flare peak and the first detection ($t_{\rm det}$). 

\begin{deluxetable*}{lrlclll} \tablewidth{0pt} \tabletypesize{\scriptsize} 
\tablecaption{Flare Model Properties \label{tab:flareprop} }
\tablehead{ \colhead{Model} & \colhead{$t_{\rm det}$}  & \colhead{$t_{1/2}$} & \colhead{Peak $F_V$} & \colhead{$\Delta V$} & \colhead{ED} & \colhead{$E_V$} \\
  \colhead{} & \colhead{(s)} & \colhead{(s)} & \colhead{(erg cm$^{-2}$ s$^{-1}$ \AA$^{-1}$)} & \colhead{(mag)} &\colhead{(s)} & \colhead{(erg)} }
\startdata
Minimum model       & 0.0 & 384.0 & 1.48$\times10^{-14}$ & $-$11.3 & 2.2$\times10^{7}$ & 4.9$\times10^{33}$\\
Best fit            & 241.0 & 178.5 & 5.19$\times10^{-14}$ & $-$12.7 & 3.7$\times10^{7}$ & 8.1$\times10^{33}$
\enddata
\end{deluxetable*}

We generated one million flare light curves based on unique combinations spanning a range of peak $F_V$, $t_{1/2}$, and $t_{det}$, then quantified their fit to the data by minimizing the square to the difference between the model and the data divided by the uncertainties. We highlight two models: a minimum model that assumes the first flare detection is the peak value (adopting  $t_{det}=0$ and the first observed value for peak $F_V$) and the model with the best overall fit in all three parameters. The properties of both models are given in Table~\ref{tab:flareprop}. The minimum model was selected to place a lower limit on the flare energy, but no strong upper limits can be placed on the flare energy because large flares often trigger sympathetic events in nearby magnetically active regions, resulting in long, multi-peaked emission events \citep[e.g.,][]{Kowalski2010,Hawley2014}. 

\subsection{Physical Properties of the Flare}
We calculate the total energy emitted in the $V$-band during the flare by integrating across the model light curves; adopting the distance of $d=97$~pc we calculate a total energy of $E_V=4.9\times10^{33}$~erg for the minimum model and $E_V=8.1\times10^{33}$~erg for the best-fit model. The quiescent $V$-band luminosity of \name\ is $L_V=2.2\times10^{26}$~erg~s$^{-1}$, indicating equivalent durations (EDs; time required to emit the flare energy during quiescence) of ED=$2.2\times10^7$~s and $3.7\times10^7$~s, respectively. The $V$-band flare energy is also significantly larger than the bolometric luminosity \citep[$\sim1.0\times10^{30}$~erg~s$^{-1}$ for spectral type L0;][]{Filippazzo2015}.

To compare this flare to those observed with Kepler on \othername, we adapt the \citet{Gizis2013} method of converting energy observed in a specific filter to total thermal energy by assuming the entirety of the flare energy is emitted in an 8000~K thermal spectrum. Due to the relatively cool temperature and the exclusion of atomic emission, this total thermal energy is a strong lower limit on the total flare energy. We calculate $E_{\rm tot}=7.7E_V$, indicating a total thermal energy of $E_{\rm tot}=3.7\times10^{34}$~erg for the minimum model and $E_{\rm tot}=6.3\times10^{34}$~erg  for the best-fit model. These values both indicate this flare was $\sim100$ times larger than the most energetic flare observed on \othername. 

Some of the largest flares on M dwarfs have been characterized in the $U$-band with $E_U\sim2\times10^{34}$~erg \citep{Hawley1991,Kowalski2010}. Adopting the \citet{Lacy1976} conversion of $E_U=1.7E_V$, we calculate $E_U\sim10^{34}$~erg, placing \name\ on par with these energetic M dwarf flares. Like those flares, however, \name\ is still less energetic than the largest stellar flares, which occur primarily on young F and G dwarfs \citep[e.g.,][]{Davenport2016}

We can also characterize the ASASSN-16ae event with a rough estimate of the size of the flare emission region. At optical wavelengths, white-light flares can be characterized by a combination of thermal (blackbody) and atomic (e.g., hydrogen Balmer) emission. The relative contribution of these two components and their characteristic temperatures change during the flare as the plasma cools. According to the detailed observations of \citet{Kowalski2013}, thermal emission (characterized by temperatures of $T=9000$--$14000$~K) contributes 95\% of the total flux during the impulsive phase, dropping to $\sim$50\% of the total flux with characteristic temperatures of $T=6000$--$10000$~K during the gradual phase. To calculate a total flare area, we consider only the thermal emission; to convert total flare area to filling factor, we adopt a radius of $R=0.105\pm0.007$~R$_{\odot}$ \citep[based on the radii of field M9--L1 dwarfs from][]{Filippazzo2015}. Due to the large flux of the flare, temperatures less than $T<7000$~K are excluded as they would cover an area larger than the stellar disk. 

If the observations occurred during the impulsive phase, the temperature range of $T=9000$--$14000$~K corresponds to total areas of 2.8--9.0$\times10^{19}$~cm$^{-2}$ (17--54\% of the surface). If the observations occurred during the gradual phase, the temperature range of $T=7000$--$10000$~K corresponds to total areas of 2.5--11$\times10^{19}$~cm$^{-2}$ (20--68\% of the surface). If the observations were instead taken during the transition from the impulsive phase to the gradual phase, as indicated by the model fit, intermediate temperatures and thermal contributions ($T=9000$--$11000$~K, $\sim$75\%) correspond to total areas of 4.0--7.1$\times10^{19}$~cm$^{-2}$ (22--40\% of the surface). Overall, the total areas are similar to the largest mid-M dwarf flares, but the filling factors for ASASSN-16ae are much larger ($>17$\% compared to $<5$\%). In contrast, the filling factors of quiescent H$\alpha$ emission drop precipitously between mid-M and early-L dwarfs \citep{Schmidt2015}. 

\section{Discussion}
\label{sec:disc}
The detection of ASASSN-16ae is the first evidence that very powerful ($E\sim10^{34}$~erg) white-light flares persist into the L spectral class and can occur on relatively old objects. If the flare frequency distribution of \othername\ \citep{Gizis2013} is typical of early-L dwarfs and can be extrapolated to higher energies, flares this large are likely to occur less frequently than on mid-M dwarfs: once every few years, compared to once or twice per month \citep{Hawley2014}. At comparable rates, ML dwarf flares could have a dramatic impact on the search for fast extragalactic transients \citep[such as kilonova gravitational-wave counterparts;][]{Cowperthwaite2015,Kasen2015} in addition to a dramatic impact on their own magnetic evolution and any exoplanet companions. 

To confirm the extrapolated L0 dwarf flare rate from \othername, we derive an order of magnitude estimate based on the ASAS-SN observations. ASAS-SN is likely to detect these powerful flares on L dwarfs as distant as 100~pc. Within that volume, we expect $\sim$5000 total L0 dwarfs \citep{Cruz2003,Cruz2007}. ASAS-SN has taken an average of 500 images (90~s each) over the entire sky, indicating a total of $2\times10^8$~s ($\sim$6~years) of L0 dwarf observations. With one flare detection, that yields a rate of one powerful flare on each L0 dwarf every six years; consistent with an extrapolation of the \othername\ flare rate. At a rate of $\sim1$ flare per $10^8$~s, magnetic energy could be dissipated by flares this size ($E\sim10^{34}$~erg) at an average rate of $10^{26}$~erg~s$^{-1}$, or roughly $10^{-4}$ times the bolometric luminosity, placing energy release through flares on par with quiescent H$\alpha$ emission and one order of magnitude smaller than the X-ray emission. 

The existence (and frequency) of dramatic flares on very small stars could also have strong a strong effect on the habitability of Earth-size planets around ML dwarfs, such as the recently discovered planetary system around M8 Trappist-1 \citep{Gillon2016}. \citet{Segura2010} found that the UV flux during large flares would be unlikely to affect habitable planets around flare star AD Leo, but habitable planets around ML dwarfs would be located $\sim$4 times closer (0.024--0.049~au compared to 0.16~au), increasing the UV flux from flares by over an order of magnitude. Persistent UV flux could also interfere with the detection of biosignatures on a habitable world \citep{Rugheimer2015}. A better understanding of magnetic activity on ML dwarfs could be essential to the detection and characterization of nearby habitable exoplanets. 

\acknowledgements
We thank the referee for useful comments that improved this Letter, and we thank Kelle L. Cruz for spectral templates, J. Davy Kirkpatrick for spectra, Jennifer van Saders for useful discussion, and LCOGT and its staff for continued support of ASAS-SN. ASAS-SN, as well as K.Z.S. and C.S.K., are supported by NSF grant AST-1515927. Development of ASAS-SN has been supported by NSF grant AST-0908816, the Center for Cosmology and AstroParticle Physics at the Ohio State University, the Mt. Cuba Astronomical Foundation, and George Skestos.

B.J.S. is supported by NASA through Hubble Fellowship grant HST-HF-51348.001 awarded by the Space Telescope Science Institute, which is operated by the Association of Universities for Research in Astronomy, Inc., for NASA, under contract NAS 5-26555. T.W.-S.H. is supported by the DOE Computational Science Graduate Fellowship, grant number DE-FG02-97ER25308.  Support for J.L.P. is in part provided by FONDECYT through the grant 1151445 and by the Ministry of Economy, Development, and Tourism's Millennium Science Initiative through grant IC120009, awarded to The Millennium Institute of Astrophysics, MAS. S.D. is supported by ``the Strategic Priority Research Program-The Emergence of Cosmological Structures'' of the Chinese Academy of Sciences (grant No. XDB09000000) and Project 11573003 supported by NSFC. J.S. acknowledges support from the Packard Foundation.

Based in part on observations obtained at the Southern Astrophysical Research (SOAR) telescope, a joint project of the Minist\'{e}rio da Ci\^{e}ncia, Tecnologia, e Inova\c{c}\~{a}o (MCTI) da Rep\'{u}blica Federativa do Brasil, the U.S. National Optical Astronomy Observatory (NOAO), the University of North Carolina at Chapel Hill (UNC), and Michigan State University (MSU).

This research benefitted from the SpeX Prism Spectral Libraries (\url{http://pono.ucsd.edu/~adam/browndwarfs/spexprism}) maintained by Adam Burgasser and the Ultracool RIZzo Spectral Library (\url{http://dx.doi.org/10.5281/zenodo.11313}) maintained by Jonathan Gagn\'e and Kelle Cruz.

This publication makes use of data products from the Two Micron All-Sky Survey, which is a joint project of the University of Massachusetts and the Infrared Processing and Analysis Center/California Institute of Technology, funded by the National Aeronautics and Space Administration and the National Science Foundation and from the \textit{Wide-field Infrared Survey Explorer}, which is a joint project of the University of California, Los Angeles, and the Jet Propulsion Laboratory/California Institute of Technology, funded by the National Aeronautics and Space Administration.

This publication also makes use of data from the Sloan Digital Sky Survey. Funding for SDSS-III has been provided by the Alfred P. Sloan Foundation, the Participating Institutions, the National Science Foundation, and the U.S. Department of Energy Office of Science. The SDSS-III web site is \url{http://www.sdss3.org/}. SDSS-III is managed by the Astrophysical Research Consortium for the Participating Institutions (listed at \url{http://www.sdss3.org/collaboration/boiler-plate.php}).



\begin{thebibliography}{}
\expandafter\ifx\csname natexlab\endcsname\relax\def\natexlab#1{#1}\fi

\bibitem[{{Ahn} {et~al.}(2014){Ahn}, {Alexandroff}, {Allende Prieto}, {Anders},
  {Anderson}, {Anderton}, {Andrews}, {Aubourg}, {Bailey}, {Bastien}, \&
  et~al.}]{Ahn2014}
{Ahn}, C.~P., {Alexandroff}, R., {Allende Prieto}, C., {et~al.} 2014, \apjs,
  211, 17

\bibitem[{{Allers} \& {Liu}(2013)}]{Allers2013}
{Allers}, K.~N., \& {Liu}, M.~C. 2013, \apj, 772, 79

\bibitem[{{Bochanski} {et~al.}(2009){Bochanski}, {Hennawi}, {Simcoe},
  {Prochaska}, {West}, {Burgasser}, {Burles}, {Bernstein}, {Williams}, \&
  {Murphy}}]{Bochanski2009}
{Bochanski}, J.~J., {Hennawi}, J.~F., {Simcoe}, R.~A., {et~al.} 2009, \pasp,
  121, 1409

\bibitem[{{Burgasser} {et~al.}(2008){Burgasser}, {Looper}, {Kirkpatrick},
  {Cruz}, \& {Swift}}]{Burgasser2008a}
{Burgasser}, A.~J., {Looper}, D.~L., {Kirkpatrick}, J.~D., {Cruz}, K.~L., \&
  {Swift}, B.~J. 2008, \apj, 674, 451

\bibitem[{{Burgasser} \& {McElwain}(2006)}]{Burgasser2006a}
{Burgasser}, A.~J., \& {McElwain}, M.~W. 2006, \aj, 131, 1007

\bibitem[{{Clemens} {et~al.}(2004){Clemens}, {Crain}, \&
  {Anderson}}]{Clemens2004}
{Clemens}, J.~C., {Crain}, J.~A., \& {Anderson}, R. 2004, \procspie,
  5492, 331--340

\bibitem[{{Cowperthwaite} \& {Berger}(2015)}]{Cowperthwaite2015}
{Cowperthwaite}, P.~S., \& {Berger}, E. 2015, \apj, 814, 25

\bibitem[{{Cruz}(2016)}]{Cruz2016}
{Cruz}, K.~L. 2016, AAS submitted

\bibitem[{{Cruz} {et~al.}(2003){Cruz}, {Reid}, {Liebert}, {Kirkpatrick}, \&
  {Lowrance}}]{Cruz2003}
{Cruz}, K.~L., {Reid}, I.~N., {Liebert}, J., {Kirkpatrick}, J.~D., \&
  {Lowrance}, P.~J. 2003, \aj, 126, 2421

\bibitem[{{Cruz} {et~al.}(2007){Cruz}, {Reid}, {Kirkpatrick}, {Burgasser},
  {Liebert}, {Solomon}, {Schmidt}, {Allen}, {Hawley}, \& {Covey}}]{Cruz2007}
{Cruz}, K.~L., {Reid}, I.~N., {Kirkpatrick}, J.~D., {et~al.} 2007, \aj, 133,
  439

\bibitem[{{Cushing} {et~al.}(2005){Cushing}, {Rayner}, \&
  {Vacca}}]{Cushing2005}
{Cushing}, M.~C., {Rayner}, J.~T., \& {Vacca}, W.~D. 2005, \apj, 623, 1115

\bibitem[{{Cushing} {et~al.}(2004){Cushing}, {Vacca}, \&
  {Rayner}}]{Cushing2004}
{Cushing}, M.~C., {Vacca}, W.~D., \& {Rayner}, J.~T. 2004, \pasp, 116, 362

\bibitem[{{Davenport}(2016)}]{Davenport2016}
{Davenport}, J.~R.~A. 2016, ArXiv e-prints, arXiv:1607.03494

\bibitem[{{Davenport} {et~al.}(2014){Davenport}, {Hawley}, {Hebb},
  {Wisniewski}, {Kowalski}, {Johnson}, {Malatesta}, {Peraza}, {Keil},
  {Silverberg}, {Jansen}, {Scheffler}, {Berdis}, {Larsen}, \&
  {Hilton}}]{Davenport2014a}
{Davenport}, J.~R.~A., {Hawley}, S.~L., {Hebb}, L., {et~al.} 2014, \apj, 797,
  122

\bibitem[{{Dieterich} {et~al.}(2014){Dieterich}, {Henry}, {Jao}, {Winters},
  {Hosey}, {Riedel}, \& {Subasavage}}]{Dieterich2014}
{Dieterich}, S.~B., {Henry}, T.~J., {Jao}, W.-C., {et~al.} 2014, \aj, 147, 94

\bibitem[{{Dupuy} \& {Liu}(2012)}]{Dupuy2012}
{Dupuy}, T.~J., \& {Liu}, M.~C. 2012, \apjs, 201, 19

\bibitem[{{Filippazzo} {et~al.}(2015){Filippazzo}, {Rice}, {Faherty}, {Cruz},
  {Van Gordon}, \& {Looper}}]{Filippazzo2015}
{Filippazzo}, J.~C., {Rice}, E.~L., {Faherty}, J., {et~al.} 2015, \apj, 810,
  158

\bibitem[{{Fuhrmeister} \& {Schmitt}(2004)}]{Fuhrmeister2004}
{Fuhrmeister}, B., \& {Schmitt}, J.~H.~M.~M. 2004, \aap, 420, 1079

\bibitem[{{Gagn{\'e}}(2016)}]{Gagne2016}
{Gagn{\'e}}, J. 2016, ApJS submitted

\bibitem[{Gagn{\'e} {et~al.}(2015)Gagn{\'e}, Lambrides, Faherty, \&
  Simcoe}]{Gagne2015x}
Gagn{\'e}, J., Lambrides, E., Faherty, J.~K., \& Simcoe, R. 2015, FireHose\_v2:
  Firehose v2.0, doi:10.5281/zenodo.18775

\bibitem[{{Gagn{\'e}} {et~al.}(2015){Gagn{\'e}}, {Faherty}, {Cruz},
  {Lafreni{\'e}re}, {Doyon}, {Malo}, {Burgasser}, {Naud}, {Artigau},
  {Bouchard}, {Gizis}, \& {Albert}}]{Gagne2015}
{Gagn{\'e}}, J., {Faherty}, J.~K., {Cruz}, K.~L., {et~al.} 2015, \apjs, 219, 33

\bibitem[{{Genzel} \& {Stutzki}(1989)}]{Genzel1989}
{Genzel}, R., \& {Stutzki}, J. 1989, \araa, 27, 41

\bibitem[{{Gillon} {et~al.}(2016){Gillon}, {Jehin}, {Lederer}, {Delrez}, {de
  Wit}, {Burdanov}, {Van Grootel}, {Burgasser}, {Triaud}, {Opitom}, {Demory},
  {Sahu}, {Bardalez Gagliuffi}, {Magain}, \& {Queloz}}]{Gillon2016}
{Gillon}, M., {Jehin}, E., {Lederer}, S.~M., {et~al.} 2016, \nat, 533, 221

\bibitem[{{Gizis} {et~al.}(2013){Gizis}, {Burgasser}, {Berger}, {Williams},
  {Vrba}, {Cruz}, \& {Metchev}}]{Gizis2013}
{Gizis}, J.~E., {Burgasser}, A.~J., {Berger}, E., {et~al.} 2013, \apj, 779, 172

\bibitem[{{Hall}(2002)}]{Hall2002}
{Hall}, P.~B. 2002, \apjl, 564, L89

\bibitem[{{Hawley} {et~al.}(2014){Hawley}, {Davenport}, {Kowalski},
  {Wisniewski}, {Hebb}, {Deitrick}, \& {Hilton}}]{Hawley2014}
{Hawley}, S.~L., {Davenport}, J.~R.~A., {Kowalski}, A.~F., {et~al.} 2014, \apj,
  797, 121

\bibitem[{{Hawley} \& {Pettersen}(1991)}]{Hawley1991}
{Hawley}, S.~L., \& {Pettersen}, B.~R. 1991, \apj, 378, 725

\bibitem[{{Henden} \& {Munari}(2014)}]{Henden2014}
{Henden}, A., \& {Munari}, U. 2014, Contributions of the Astronomical
  Observatory Skalnate Pleso, 43, 518

\bibitem[{{Hilton}(2011)}]{Hilton2011phd}
{Hilton}, E.~J. 2011, PhD thesis, University of Washington

\bibitem[{{Kasen} {et~al.}(2015){Kasen}, {Fern{\'a}ndez}, \&
  {Metzger}}]{Kasen2015}
{Kasen}, D., {Fern{\'a}ndez}, R., \& {Metzger}, B.~D. 2015, \mnras, 450, 1777

\bibitem[{{Kirkpatrick} {et~al.}(1999){Kirkpatrick}, {Reid}, {Liebert},
  {Cutri}, {Nelson}, {Beichman}, {Dahn}, {Monet}, {Gizis}, \&
  {Skrutskie}}]{Kirkpatrick1999}
{Kirkpatrick}, J.~D., {Reid}, I.~N., {Liebert}, J., {et~al.} 1999, \apj, 519,
  802

\bibitem[{{Kirkpatrick} {et~al.}(2010){Kirkpatrick}, {Looper}, {Burgasser},
  {Schurr}, {Cutri}, {Cushing}, {Cruz}, {Sweet}, {Knapp}, {Barman},
  {Bochanski}, {Roellig}, {McLean}, {McGovern}, \& {Rice}}]{Kirkpatrick2010}
{Kirkpatrick}, J.~D., {Looper}, D.~L., {Burgasser}, A.~J., {et~al.} 2010,
  \apjs, 190, 100

\bibitem[{{Kirkpatrick} {et~al.}(2014){Kirkpatrick}, {Schneider},
  {Fajardo-Acosta}, {Gelino}, {Mace}, {Wright}, {Logsdon}, {McLean}, {Cushing},
  {Skrutskie}, {Eisenhardt}, {Stern}, {Balokovi{\'c}}, {Burgasser}, {Faherty},
  {Lansbury}, {Rich}, {Skrzypek}, {Fowler}, {Cutri}, {Masci}, {Conrow},
  {Grillmair}, {McCallon}, {Beichman}, \& {Marsh}}]{Kirkpatrick2014}
{Kirkpatrick}, J.~D., {Schneider}, A., {Fajardo-Acosta}, S., {et~al.} 2014,
  \apj, 783, 122

\bibitem[{{Kowalski} {et~al.}(2010){Kowalski}, {Hawley}, {Holtzman},
  {Wisniewski}, \& {Hilton}}]{Kowalski2010}
{Kowalski}, A.~F., {Hawley}, S.~L., {Holtzman}, J.~A., {Wisniewski}, J.~P., \&
  {Hilton}, E.~J. 2010, \apjl, 714, L98

\bibitem[{{Kowalski} {et~al.}(2013){Kowalski}, {Hawley}, {Wisniewski}, {Osten},
  {Hilton}, {Holtzman}, {Schmidt}, \& {Davenport}}]{Kowalski2013}
{Kowalski}, A.~F., {Hawley}, S.~L., {Wisniewski}, J.~P., {et~al.} 2013, \apjs,
  207, 15

\bibitem[{{Lacy} {et~al.}(1976){Lacy}, {Moffett}, \& {Evans}}]{Lacy1976}
{Lacy}, C.~H., {Moffett}, T.~J., \& {Evans}, D.~S. 1976, \apjs, 30, 85

\bibitem[{{Liebert} {et~al.}(2003){Liebert}, {Kirkpatrick}, {Cruz}, {Reid},
  {Burgasser}, {Tinney}, \& {Gizis}}]{Liebert2003}
{Liebert}, J., {Kirkpatrick}, J.~D., {Cruz}, K.~L., {et~al.} 2003, \aj, 125,
  343

\bibitem[{{Liebert} {et~al.}(1999){Liebert}, {Kirkpatrick}, {Reid}, \&
  {Fisher}}]{Liebert1999}
{Liebert}, J., {Kirkpatrick}, J.~D., {Reid}, I.~N., \& {Fisher}, M.~D. 1999,
  \apj, 519, 345

\bibitem[{{Marshall} {et~al.}(2008){Marshall}, {Burles}, {Thompson},
  {Shectman}, {Bigelow}, {Burley}, {Birk}, {Estrada}, {Jones}, {Smith},
  {Kowal}, {Castillo}, {Storts}, \& {Ortiz}}]{Marshall2008}
{Marshall}, J.~L., {Burles}, S., {Thompson}, I.~B., {et~al.} 2008, 
  \procspie, 7014, 701454

\bibitem[{{Munn} {et~al.}(2004){Munn}, {Monet}, {Levine}, {Canzian}, {Pier},
  {Harris}, {Lupton}, {Ivezi{\'c}}, {Hindsley}, {Hennessy}, {Schneider}, \&
  {Brinkmann}}]{Munn2004}
{Munn}, J.~A., {Monet}, D.~G., {Levine}, S.~E., {et~al.} 2004, \aj, 127, 3034

\bibitem[{{Pineda} {et~al.}(2016){Pineda}, {Hallinan}, {Kirkpatrick}, {Cotter},
  {Kao}, \& {Mooley}}]{Pineda2016}
{Pineda}, J.~S., {Hallinan}, G., {Kirkpatrick}, J.~D., {et~al.} 2016, ArXiv
  e-prints, arXiv:1604.03941

\bibitem[{{Reid} {et~al.}(2008){Reid}, {Cruz}, {Kirkpatrick}, {Allen},
  {Mungall}, {Liebert}, {Lowrance}, \& {Sweet}}]{Reid2008}
{Reid}, I.~N., {Cruz}, K.~L., {Kirkpatrick}, J.~D., {et~al.} 2008, \aj, 136,
  1290

\bibitem[{{Rockenfeller} {et~al.}(2006){Rockenfeller}, {Bailer-Jones}, {Mundt},
  \& {Ibrahimov}}]{Rockenfeller2006}
{Rockenfeller}, B., {Bailer-Jones}, C.~A.~L., {Mundt}, R., \& {Ibrahimov},
  M.~A. 2006, \mnras, 367, 407

\bibitem[{{Rodr{\'{\i}}guez-Barrera} {et~al.}(2015){Rodr{\'{\i}}guez-Barrera},
  {Helling}, {Stark}, \& {Rice}}]{Rodriguez-Barrera2015}
{Rodr{\'{\i}}guez-Barrera}, M.~I., {Helling}, C., {Stark}, C.~R., \& {Rice},
  A.~M. 2015, \mnras, 454, 3977

\bibitem[{{Rugheimer} {et~al.}(2015){Rugheimer}, {Kaltenegger}, {Segura},
  {Linsky}, \& {Mohanty}}]{Rugheimer2015}
{Rugheimer}, S., {Kaltenegger}, L., {Segura}, A., {Linsky}, J., \& {Mohanty},
  S. 2015, \apj, 809, 57

\bibitem[{{Schmidt}(2016)}]{Schmidt2016b}
{Schmidt}, S.~J. 2016, in prep.

\bibitem[{{Schmidt} {et~al.}(2007){Schmidt}, {Cruz}, {Bongiorno}, {Liebert}, \&
  {Reid}}]{Schmidt2007}
{Schmidt}, S.~J., {Cruz}, K.~L., {Bongiorno}, B.~J., {Liebert}, J., \& {Reid},
  I.~N. 2007, \aj, 133, 2258

\bibitem[{{Schmidt} {et~al.}(2015){Schmidt}, {Hawley}, {West}, {Bochanski},
  {Davenport}, {Ge}, \& {Schneider}}]{Schmidt2015}
{Schmidt}, S.~J., {Hawley}, S.~L., {West}, A.~A., {et~al.} 2015, \aj, 149, 158

\bibitem[{{Schmidt} {et~al.}(2014{\natexlab{a}}){Schmidt}, {West}, {Bochanski},
  {Hawley}, \& {Kielty}}]{Schmidt2014a}
{Schmidt}, S.~J., {West}, A.~A., {Bochanski}, J.~J., {Hawley}, S.~L., \&
  {Kielty}, C. 2014{\natexlab{a}}, \pasp, 126, 642

\bibitem[{{Schmidt} {et~al.}(2010){Schmidt}, {West}, {Hawley}, \&
  {Pineda}}]{Schmidt2010}
{Schmidt}, S.~J., {West}, A.~A., {Hawley}, S.~L., \& {Pineda}, J.~S. 2010, \aj,
  139, 1808

\bibitem[{{Schmidt} {et~al.}(2014{\natexlab{b}}){Schmidt}, {Prieto}, {Stanek},
  {Shappee}, {Morrell}, {Bardalez Gagliuffi}, {Kochanek}, {Jencson}, {Holoien},
  {Basu}, {Beacom}, {Szczygie{\l}}, {Pojmanski}, {Brimacombe}, {Dubberley},
  {Elphick}, {Foale}, {Hawkins}, {Mullins}, {Rosing}, {Ross}, \&
  {Walker}}]{Schmidt2014}
{Schmidt}, S.~J., {Prieto}, J.~L., {Stanek}, K.~Z., {et~al.}
  2014{\natexlab{b}}, \apjl, 781, L24

\bibitem[{{Segura} {et~al.}(2010){Segura}, {Walkowicz}, {Meadows}, {Kasting},
  \& {Hawley}}]{Segura2010}
{Segura}, A., {Walkowicz}, L.~M., {Meadows}, V., {Kasting}, J., \& {Hawley}, S.
  2010, Astrobiology, 10, 751

\bibitem[{{Shappee}(2016)}]{Shappee2016b}
{Shappee}, B.~J. 2016, in prep.

\bibitem[{{Shappee} {et~al.}(2014){Shappee}, {Prieto}, {Grupe}, {Kochanek},
  {Stanek}, {De Rosa}, {Mathur}, {Zu}, {Peterson}, {Pogge}, {Komossa}, {Im},
  {Jencson}, {Holoien}, {Basu}, {Beacom}, {Szczygie{\l}}, {Brimacombe},
  {Adams}, {Campillay}, {Choi}, {Contreras}, {Dietrich}, {Dubberley},
  {Elphick}, {Foale}, {Giustini}, {Gonzalez}, {Hawkins}, {Howell}, {Hsiao},
  {Koss}, {Leighly}, {Morrell}, {Mudd}, {Mullins}, {Nugent}, {Parrent},
  {Phillips}, {Pojmanski}, {Rosing}, {Ross}, {Sand}, {Terndrup}, {Valenti},
  {Walker}, \& {Yoon}}]{Shappee2014}
{Shappee}, B.~J., {Prieto}, J.~L., {Grupe}, D., {et~al.} 2014, \apj, 788, 48

\bibitem[{{Shappee} {et~al.}(2016){Shappee}, {Stanek}, {Schmidt}, {Holoien},
  {Brown}, {Kochanek}, {Godoy-Rivera}, {Basu}, {Prieto}, {Bersier}, {Dong},
  {Chen}, \& {Brimacombe}}]{Shappee2016}
{Shappee}, B.~J., {Stanek}, K.~Z., {Schmidt}, S., {et~al.} 2016, The
  Astronomer's Telegram, 8553

\bibitem[{{Simcoe} {et~al.}(2013){Simcoe}, {Burgasser}, {Schechter}, {Fishner},
  {Bernstein}, {Bigelow}, {Pipher}, {Forrest}, {McMurtry}, {Smith}, \&
  {Bochanski}}]{Simcoe2013}
{Simcoe}, R.~A., {Burgasser}, A.~J., {Schechter}, P.~L., {et~al.} 2013, \pasp,
  125, 270

\bibitem[{{Skrutskie} {et~al.}(2006){Skrutskie}, {Cutri}, {Stiening},
  {Weinberg}, {Schneider}, {Carpenter}, {Beichman}, {Capps}, {Chester},
  {Elias}, {Huchra}, {Liebert}, {Lonsdale}, {Monet}, {Price}, {Seitzer},
  {Jarrett}, {Kirkpatrick}, {Gizis}, {Howard}, {Evans}, {Fowler}, {Fullmer},
  {Hurt}, {Light}, {Kopan}, {Marsh}, {McCallon}, {Tam}, {Van Dyk}, \&
  {Wheelock}}]{Skrutskie2006}
{Skrutskie}, M.~F., {Cutri}, R.~M., {Stiening}, R., {et~al.} 2006, \aj, 131,
  1163

\bibitem[{{Stelzer} {et~al.}(2006){Stelzer}, {Schmitt}, {Micela}, \&
  {Liefke}}]{Stelzer2006}
{Stelzer}, B., {Schmitt}, J.~H.~M.~M., {Micela}, G., \& {Liefke}, C. 2006,
  \aap, 460, L35

\bibitem[{{Vacca} {et~al.}(2003){Vacca}, {Cushing}, \& {Rayner}}]{Vacca2003}
{Vacca}, W.~D., {Cushing}, M.~C., \& {Rayner}, J.~T. 2003, \pasp, 115, 389

\bibitem[{{Wright} {et~al.}(2010){Wright}, {Eisenhardt}, {Mainzer}, {Ressler},
  {Cutri}, {Jarrett}, {Kirkpatrick}, {Padgett}, {McMillan}, {Skrutskie},
  {Stanford}, {Cohen}, {Walker}, {Mather}, {Leisawitz}, {Gautier}, {McLean},
  {Benford}, {Lonsdale}, {Blain}, {Mendez}, {Irace}, {Duval}, {Liu}, {Royer},
  {Heinrichsen}, {Howard}, {Shannon}, {Kendall}, {Walsh}, {Larsen}, {Cardon},
  {Schick}, {Schwalm}, {Abid}, {Fabinsky}, {Naes}, \& {Tsai}}]{Wright2010}
{Wright}, E.~L., {Eisenhardt}, P.~R.~M., {Mainzer}, A.~K., {et~al.} 2010, \aj,
  140, 1868

\end{thebibliography}
\end{document}